\documentstyle[12pt]{article}
\begin{document}
\large
\begin{center}
\bf{Are Pressure-Confined Clouds in Galactic Halo\\
Possible for a Model of Ly$\alpha$ Clouds?}
\vspace{2cm}\\
\normalsize
\rm
    Keiko {\sc Miyahata}
\footnote{*JSPS junior fellow}
 and Satoru {\sc Ikeuchi}
\footnote{*present address; Department of Physics, Nagoya University, 
Furo-cho, Chikusa-ku, Nagoya 464-01, 
ikeuchi@a.phys.nagoya-u.ac.jp}
\\
\vspace{1cm}
\it
 Department of Earth and Space Science, Faculty of Science,
 Osaka University,\\
 Machikaneyama-cho, Toyonaka, Osaka 560
\vspace{12pt}\\
 E-mail miyahata@vega.ess.sci.osaka-u.ac.jp,
ikeuchi@vega.ess.sci.osaka-u.ac.jp
\vspace{1.0cm}\\
\end{center}
\baselineskip 12mm
\sloppy{
\begin{center}
\large
\bf{Abstract} \\
\end{center}

\normalsize

\baselineskip 12mm
Our understanding of the 
Lyman $\alpha$ forest 
has considerably changed between 
before and after the {\it{Hubble Space Telescope}} and Keck Telescope
 in operation, 
 because the Ly$\alpha$ clouds at low redshifts (z$<$1.7)  
observed by {\it{HST}} showed us two unexpected features: 
Lanzetta et al (1995) found that 
most luminous 
galaxies at such redshifts produce Ly$\alpha$ absorptions at 
the mean impact parameter $\sim$160$h^{-1}$kpc and establised 
the assosiation between Ly$\alpha$ clouds and galaxies. 
Ulmer (1996) pointed out the strong clustering of Ly$\alpha$ clouds 
in this redshift range. 
Motivated by these observations 
together with another observation which reports the detection of 
metal in the Ly$\alpha$ clouds at high redshift universe, 
 we propose the two-component protogalaxy model
as a model for the Ly$\alpha$ cloud
 based upon the previous work (Miyahata,Ikeuchi 1995). 
In our model, the Ly$\alpha$ clouds 
are supposed to be stable cold clouds confined 
by the pressure of ambient hot gas in galactic halo.
We determine the properties of these cold clouds and hot gas 
on the basis of theoretical and observational constraints.
 Especially, we take into account the stability 
of a cold cloud in the galactic halo 
in addition to the general stability 
conditions in a two-component medium which was discussed 
in Ikeuchi,Ostriker(1986), and compare the derived quantities of Ly$\alpha$ 
clouds both cases in galactic halo and in intergalactic medium between 
high and low redshifts.
 We conclude that the condition that a cloud is stable against both 
evaporation and tidal destruction by a hot galactic halo is very restrictive.
As a result, in the most noteworthy case, at z${\sim}$0.5,
it is concluded that 
 a pressure-confined, stable spherical
 Ly$\alpha$ cloud with typical neutral hydrogen column 
density $N_{HI}=10^{14}cm^{-2}$
 cannot survive in the galactic halo, although much higher column density 
clouds of $N_{HI}=10^{17}cm^{-2}$ can do there. 
 We discuss how our result constrains an alternative model for Ly$\alpha$ 
 clouds which associate with galaxy observed by Lanzetta et.al.\\
\bf{Key words}
\rm
:~Galaxies:~evolution---halo---ISM---Quasars:~absorption lines
\vspace{1.5cm}\\
\large
\bf{1.~Introduction}
\rm
\normalsize
\baselineskip 12mm

~~The {\it{Hubble Space Telescope}} and Keck Telescope have revealed 
the history of the universe from high redshifts to 
the present epoch, and our
knowledge of galaxy formation and evolution has made a remarkable 
progress
 (Fukugita, et.al.1996).
 In this context, 
our understanding
 of the Ly$\alpha$ cloud has considerably changed before and after 
the {\it{HST}} and Keck Telescope 
in operation (Ikeuchi 1995; Shull 1997).
 Before these telescopes, 
we thought that the Ly$\alpha$ clouds are the 
intergalactic primordial gas clouds which are confined by the 
thermal pressure of intergalactic medium  
(Ostriker, Ikeuchi 1983; Ikeuchi, Ostriker 1986)
and/or by the gravity of the cold dark matter
(Ikeuchi 1986, Rees 1986).
Therefore, it was supposed that they distribute randomly and
 have no or little 
correlations with galaxies.
 But a lot of new observational results have brought to us, which are not 
explained by the above standard picture of the Ly$\alpha$ clouds.
Some of them suggest that remarkable fraction of the Ly$\alpha$ clouds
 have their origin in storong connection with galaxy.

 First, for nearby ($z<1.7$) Ly$\alpha$ clouds, 
it enables us for the first time to study whether 
they associate with galaxies or not by using {\it{HST}}.
(Lanzetta et.al. 1995; Bowen et.al. 1996; Brun et.al. 1996). 
 For example, Lanzetta et.al. reported that
 (1) at redshift $<1$ most galaxies are surrounded by extended gaseous 
envelopes of its extention $\sim$160$h^{-1}$kpc with the covering factor 
roughly unity, 
and (2) many or most Ly$\alpha$ absorption systems arise in 
extended gaseous envelopes of galaxies.
 Most recent data strengthen this previous result 
(Chen et.al. 1997; Lanzetta et.al. 1997a, 1997b), 
 although the conclusion is now controversial 
(Bowen et.al. 1996; Brun et.al. 1996; Morris 1996). 
 Second, the strong clustering of Ly$\alpha$ clouds was discovered 
 in this redshift(Ulmer 1996).
 Third, recent observations by Keck telescope indicate that CIV lines are 
generally associated with the Ly$\alpha$ clouds down to moderately low column 
densities($N_{HI}{\sim}10^{14.5}cm^{-2}$) even in high redshift 
(Tytler 1995; Cowie, Songaila 1995).

Theoretically,  there are 
few models in which  clouds in the galactic halo are 
 the possible origin of Ly$\alpha$ absorption 
since Bahcall and Spitzer (1969), who proposed that extended gaseous galactic 
 halo causes absorption lines in QSO spectra.
On the othe hand, 
 Weymann(1995) emphasised that at least two populations of Ly$\alpha$ clouds
 are needed
 to explain many important characteristics about them reported until now.
 One is relatively unclustered, rapidly evolving population which 
dominates at high redshifts, and the other is more stable, possibly 
associated with galaxies which dominate at low redshifts. 
 
As a probable model for the latter population of Ly$\alpha$ clouds, 
we proposed 
two-component protogalaxy model (Miyahata, Ikeuchi 1995).
 This model was originally proposed in the context of the globular 
cluster formation (Fall, Rees 1985, Murray, Lin 1993, and references therein),
 and have been investigated in many context of galaxy formation and evolution
(Ferrara, Field 1994; Ikeuchi, Norman 1991; Norman 1994; Spaans, Norman 1997; 
Field et.al. 1997 and references therein).
 When a protogalaxy collapses and virializes to its radius 
$\sim$100kpc (Rees,Ostriker 1977), small density 
fluctuations grow up in its halo due to the thermal instability
(Field 1965).
 As a result, cold (${\sim}10^{4}$K) and small (${\sim}10^{6}M_{\odot}$) 
clouds are formed in protogalactic halo (Fall, Rees 1985).
 Here, we analyze these cold clouds as the 
candidate of Ly$\alpha$ clouds.
It is noted that in our model, the theoretically predicted scale of 
protogalactic halo is around
$\sim$100kpc which is just the same order of impact parameter 
indicated by Lanzetta et.al.(1995,1997).
By now, direct confirmation of the existence of 
hot halo extended to $>$100kpc 
has not been reported although the hot extended X-ray
emitting gas is generally observed in the elliptical galaxies out to 
$\sim$50kpc (Sarazin 1996), the highly ionized gas is obsereved far away 
from the disk of spiral galaxies in our Galaxy (Spitzer 1990),
 and the probability of metal-line absorption of QSO suggests such an 
extended gaseous galactic halo (Spitzer, Ostriker 1997).

From the viewpoint of above-mentioned two-population model for the 
Ly$\alpha$ clouds, Chiba, Nath (1997) discussed the origin of metallicity 
in high redshift Ly$\alpha$ clouds. 
They showed that the fraction of Ly$\alpha$ lines with associated 
metal lines can be understood in terms of the Ly$\alpha$ 
absorbers assoiated with galactic halo, assuming that 
pressure confined Ly$\alpha$-absorbing clouds are embedded in galactic halo.
Recently, several large numerical simulations of the large scale structure 
formation in the universe were done to examine the origin and evolution 
of the Ly$\alpha$ clouds 
(Miralda-Escude, et.al. 1996; Cen, Simcoe 1997; 
Zhang et.al. 1997; Weinberg et.al. 1997 and references therein).
The purpose of these simulations seems to explain the characteristics of the 
former population of the Ly$\alpha$ cloud which Weymann noted (1995), but 
even by such high resolution simulations it is impossible to resolve the 
galactic scale.
 So it would be emphasized here that it is important to analyze  
{\it{whether or not the Ly$\alpha$ 
clouds can exist near or inside the galactic halo}}
 by using a simple analytical model.

In this paper,  we determine the physical quantities of cold clouds which 
are confined by the pressure of ambient hot galactic halo gas 
on the basis of theoretical and observational constraints in detail.
 In section 2, 
we discuss the stability conditions for both of a cold cloud and general hot 
ambient medium and compare the results 
with the observations between z=2.5 and z=0.5. 
These cold clouds can be applied to the intergalactic Ly$\alpha$ cloud.
 In section 3, in addition to the above discussion, we examine the stability 
of a small cloud in galactic halo and compare them
 with those of an intergalactic cloud. 
From the stability analyses of the calculated cloud model, we conclude that 
a pressure-confined, stable spherical Ly$\alpha$ cloud with typical neutral 
hydrogen column density $N_{HI}=10^{14}cm^{-2}$ cannot survive in galactic 
halo for both cases of z=2.5 and z=0.5, although higher column density systems 
such as $N_{HI}=10^{17}cm^{-2}$ can do there.
 In Section 4, we discuss several points which might affect our 
conclusion, and speculate an alternative model for such 
Ly$\alpha$ clouds that Lanzetta et.al. observed. 
\vspace{1cm}\\
\large
\bf{2. Pressure Confined Clouds in Two-Component Intergalactic Medium}\\
\normalsize
\it
{2.1. Basic Equations and Assumptions}
\rm
\baselineskip 12mm

Following the discussion by Ostriker, Ikeuchi(1983) and Ikeuchi, 
Ostriker(1986),
 we assume that the cold cloud is spherical and homogeneous
 for simplicity. Suppose that 
the cold cloud is embedded in a hot ambient gas.
 In our simple treatment, no dark matter is considered.

First, we assume pressure equilibrium between 
the cold cloud and the hot ambient medium as,
\begin{eqnarray}
\tilde{P}=n_{c}T_{c}=n_{h}T_{h}.
\label{eqn:pc}
\end{eqnarray}
The notations used above and hereafter are summerized in Table 1.
Irradiated by the UV background radiation, 
the cold cloud is thought to be ionized, 
in thermal and ionization equilibria.
From the latter assumption, the following relation holds;
\begin{eqnarray}
\Gamma_{H}n_{HI}=\alpha_{H}n_{HII}n_{e},
\label{eqn:ione}
\end{eqnarray}
where  we adopt the recombination coefficient 
$\alpha_{H}=4.36{\times}10^{-10}T^{-3/4}$ at T$>5000$K and 
 $\Gamma_{H}$
 is the ionization rate of neutral hydrogen and is written as (Black 1981)
\begin{eqnarray}
\Gamma_{H}=J(\nu_{T})G_{H}.
\end{eqnarray}
Here, $\Gamma_{H}$ is the ionization rate of neutral hydrogen 
 and $G_{H}={\int}^{\nu_{max}}_{\nu_{T}}({\nu}/{\nu_{T}})^{-1}
{\sigma}({\nu})d{\nu}$ where $\nu_{T}$ and ${\sigma}({\nu})$ 
 are the frequency at Lyman Limit of hydrogen 
and the cross section for photoionization, 
 respectively.

Coupling eq.(\ref{eqn:ione}) with an assumption of thermal equilibrium for 
a cloud, Ikeuchi, Ostriker (1986) showed that the equilibrium temperature 
$T_{c}$ is always 3${\times}{\times}10^{4}$K whenever the pressure-confined 
cloud is 
embedded in an expanding IGM and the cloud is almost fully ionized. So,  
\begin{eqnarray}
n_{HII}{\sim}n_{e}{\sim}n_{c}.
\label{eqn:fion}
\end{eqnarray}
The order of $T_{c}$ is determined by the fact that the  
cooling function 
has a sharp cut-off at this temperature because of 
recombination to hydrogen atoms (Binney,Tremaine 1987).  
 Recent observations indicate that CIV lines are generally 
associated with the Ly$\alpha$ 
clouds in high redshifts (Tytler 1995, Cowie et.al. 1995). 
These are very important in relation to their origin, but such a 
low metalicity as Z${\sim}1/100Z_{\odot}$ does not affect our results at all.

Under the above assumptions, we can derive the radius and mass of the cold 
cloud as follows:
\begin{eqnarray}
R_{c}
&=&1.85{\times}10^{3}J_{-21}T_{4}^{11/4}N_{14}\tilde{P}^{-2}~~pc,
\label{eqn:rc}\\
M_{c}
&=&6.63{\times}10^{4}J_{-21}^{3}T_{4}^{29/4}N_{14}^{3}\tilde{P}^{-5}~~
M_{\odot},
\label{eqn:mc}
\end{eqnarray}
where $J_{-21}=J/(10^{-21}erg~cm^{-2}s^{-1}sr^{-1}Hz^{-1})$, 
$T_{4}=T_{c}/10^{4}K$ and $N_{14}=N_{HI}/10^{14}cm^{-2}$, 
$N_{HI}$ being the HI column density. 
It is defined as 
\begin{eqnarray}
N_{HI}=n_{HI}R_{c},
\label{eqn:nh1}
\end{eqnarray}
where we neglect the geometrical factor. 
Using eq.(\ref{eqn:rc}), we can rewrite 
eq. (\ref{eqn:nh1}) as follows,
\begin{eqnarray}
N_{HI}&=&4.1{\times}10^{13}(\frac{R_{c}}{10kpc})\tilde{P}^{2}
T_{4}^{-11/4}J_{-21}^{-1}~~~cm^{-2},
\label{eqn:nh1r}\\
&=&1.45{\times}10^{12}(\frac{M_{c}}{10^8M_{\odot}})^{1/3}\tilde{P}^{5/3}
T_{4}^{-29/12}J_{-21}^{-1}~~~cm^{-2}.
\label{eqn:nh1m}
\end{eqnarray}
Note that the neutral hydrogen column density which is observed directly 
is sensitive to $\tilde{P}(=n_{c}T_{c})$ but not sensitive to $M_{c}$.
\vspace{0.5cm}\\
\it
{2.2. Stability Conditions in Two-component System}\\
\rm
\baselineskip 12mm

~~In this subsection 
we examine the stability conditions in a two-phase medium,
and give a constraint to the physical quantities of the cold cloud 
which we derive in the previous subsection.\\
(i)The ambient medium must keep hot, 
otherwise the two-component system does not maintain any longer.
 Thus, we get the constraint,
\begin{eqnarray}
\tau_{cool}>\tau_{H},
\label{eqn:tcth}
\end{eqnarray}
where $\tau_{cool}$ is the cooling time for hot ambient gas  
and $\tau_{H}$ is the age of universe as
\begin{eqnarray}
\tau_{cool}=1.5\frac{kT_{h}}{n_{h}\Lambda{(T_{h})}},
~~\tau_{H}=\frac{4.1{\times}10^{17}}{(1+z)^{1.5}}
\hspace{0.5cm}sec,
\end{eqnarray}
where $\Lambda$ is the cooling function of primordial gas 
in an absence of external UV radiation 
(Binney, Tremaine 1987). 
In fact, the effective cooling rate changes in the presence of 
UV background radiation (Thoul, Weinberg 1996). 
But as we show later, $T_{h}$ is typically higher than ${\sim}10^{5}$K 
and our treatment is valid in this temperature range.
Here, we assume $\Omega=1$ and $H_{0}=50kms^{-1}Mpc{-1}$ for simplicity.
For these parameters, eq.(\ref{eqn:tcth}) is rewritten as follows;
\begin{eqnarray}
T_{h}>10^{33.3}n_{h}\Lambda(T_{h})(1+z)^{-1.5}~~K.
\label{eqn:s1}
\end{eqnarray}
 This condition becomes more severe in the situation 
with larger $n_{h}$.\\
(ii)The clouds must be gravitationally stable,
 othewise 
we can no longer observe such clouds as Ly $\alpha$ absorbers 
due to rapid collapse.
This implies
\begin{eqnarray}
M_{c}<M_{Jeans,p},
\label{eqn:jeans}
\end{eqnarray}
where the Jeans mass of a pressure confined cloud $M_{Jeans,p}$ 
is given by (Spitzer 1978)
\begin{eqnarray}
M_{Jeans,p}=1.18\frac{(kT_{c}/m_{p})^{2}}{G^{1.5}(k\tilde{P})^{0.5}}.
\end{eqnarray}
Using eq.(\ref{eqn:mc}), it is shown that this condition 
gives a lower bound on $\tilde{P}$, i.e.,
\begin{eqnarray}
\tilde{P}>10^{-0.996}(N_{14}J_{-21})^{2/3}T_{4}^{7/6}~~cm^{-3}K.
\label{eqn:s2}
\end{eqnarray}
(iii)
The cold clouds must not be evaporated by the heat conduction 
from ambient hot medium.
This implies
\begin{eqnarray}
\tau_{evap}>\tau_{H},
\label{eqn:s3-o}
\end{eqnarray}
where $\tau_{evap}$ is the evaporation timescale. 
 When we estimate this timescale, we must examine the saturation parameter 
$\delta=1.2{\times}10^{4}T_{h}^{2}n_{h}^{-1}R_{c}^{-1}$ 
 (Balbus, McKee 1982) which is essentially the ratio of the 
mean free path of an electron to the scale of cooled region. 
 For $\delta{\leq}$1, in so called classical case, 
the evaporation time is expressed by
\begin{eqnarray}
\tau_{evp}=1.1{\times}10^{-9}(R_{c}n_{c})^{2}T_{c}T_{h}^{-3.5}n_{h}^{-1}
\hspace{1cm}sec.
\end{eqnarray}
Using eqs. (\ref{eqn:pc}) and (\ref{eqn:rc}), this condition (\ref{eqn:s3-o}) 
is rewritten as follows,
\begin{eqnarray}
T_{h}<10^{2.35}n_{h}^{-0.545}(N_{14}J_{-21})^{0.364}T_{4}^{0.818}(1+z)^{0.237}
~~K.
\label{eqn:s3}
\end{eqnarray}
As we show in the next section,
 this simple estimate can be applied to a considerable 
 range of physical quantities of two-component protogalaxy. \\
 For 1${\leq}{\delta}{\leq}10^{2}$, when the evaporation is saturated, 
$\tau_{evp}$ is expressed as 
\begin{eqnarray}
\tau_{evp}=3.2{\times}10^{-6}N_{c}^{7/6}T_{c}^{1/6}(n_{h}T_{h})^{-1}
\hspace{1cm}sec,
\end{eqnarray}
and in this case, an upper bound on $\tilde{P}$ is given by,
\begin{eqnarray}
\tilde{P}<10^{-0.794}(N_{14}J_{-21})^{0.538}T_{4}^{1.02}(1+z)^{0.692}
~~cm^{-3}K.
\end{eqnarray}
 For ${\delta}{\geq}10^{2}$, the suprathermal evaporation must be 
 examined and in this case $\tau_{evp}$ is as follows;
\begin{eqnarray}
\tau_{evp}=1.1{\times}10^{-5}R_{c}^{4/3}n_{h}^{1/3}T_{h}^{-1/6}T_{c}^{-1}
\hspace{1cm}sec.
\end{eqnarray}
In this case, eq.(\ref{eqn:s3}) is rewritten as follows;
\begin{eqnarray}
T_{h}<10^{0.549}n_{h}^{-0.824}(N_{14}J_{-21})^{0.471}
T_{4}^{0.941}(1+z)^{0.529}~~K.
\end{eqnarray}
In any case, if the density or temperature of the ambient hot gas is high 
enough, or the cold cloud is small, the cloud will be evaporated quickly.

Using these three stability conditions, 
we can give a constraint to the cloud quantities such as 
$R_{c}$(eq.(\ref{eqn:rc})), $M_{c}$(eq.(\ref{eqn:mc}))
and $N_{HI}$(eq.(\ref{eqn:nh1})) 
through constraints given to ($n_{h},T_{h}$) or ($n_{c},T_{c}$).\\
\it
{2.3. Results}
\rm

In Figure 1a, we show three stability conditions plotted in  
the density and temperature of hot phase gas for z=2.5 and in this redshift
$J_{-21}$ is estimated nearly unity (Savaglio, Webb 1995).
For other parameters, we summerize them in Table 3 (Model 1). 
This figure reproduces the essence of stability conditions for 
the high redshift, intergalactic pressure-confined Ly$\alpha$ clouds
(Ostriker, Ikeuchi 1983).
 As is seen, the right below, left below and right upper 
region in this figure are forbidden due to the 
 cooling condition, gravitationally unstable condition, 
and evaporation condition, respectively.  
 For comparison, a constant cloud radius $R_{c}$(see eq.(\ref{eqn:rc})) 
(short-dashed line) and a constant cloud 
mass $M_{c}$(see eq.(\ref{eqn:mc}))(dot-dashed line) are shown.
 As a result, the allowed region for the pressure-confined 
clouds to survive in two-phase media is in the density region 
between $10^{-7}$ and $10^{-5}cm^{-3}$ and temperature region 
between $10^{5.5}$ and $10^{6.7}$K.
 For the median value of $n_{h}{\sim}10^{-6}cm^{-3}$, the 
typical cloud radius is $\sim$50kpc and mass $\sim10^7~M_{\odot}$, 
which is comparable to the lower bound of the 
observed size of high redshift Ly$\alpha$ absorbers 
(Bechtold,et.al 1994; Dinshaw et.al. 1994).
These results and those in hereafter are summerized in Table 4.

In Figure 1b, we show the allowed range of the cloud mass with respect to 
the ambient pressure. 
 The upper and lower mass bounds of the cloud are 
constrained by the stability condition against the Jeans instability 
and evaporation, respectively. 
The allowed region is limited to the trapezoid region left below for  
three cases of the density of the hot halo(solid and short-dashed line). 
For comparison, a constant cloud radius $R_{c}$(\ref{eqn:rc})
(short-dashed line)
 and a constant neutral hydrogen column 
density of the cloud $N_{HI}$(\ref{eqn:nh1r})
(dotted line) are also shown.
 From this figure, we can see that the wide range of the observed
 column density of Ly$\alpha$ cloud is reproduced in our model.

 Figure 2a is the same as in Figure 1a, but for z=0.5 (Model 2).
 Since the UV background radiation dramatically drops off from high redshift 
to the present (Savaglio, Webb 1994), 
we assume the parameter as $J_{-21}$=0.01.
 Figure 2b is equivalent to Figure 1b but for z=0.5.

The cooling and evaporation condition are more stringent at z=0.5  
compared to those at z=2.5. 
On the other hand, because the UV background radiation 
dramatically decreases from z$\sim$2.5 to z=0.5, the cloud radius and mass 
become smaller for the same value of $\tilde{P}$ and $N_{HI}$.  
 As a result, the allowed region for ambient hot gas exists in the 
density range  from
$n_{h}{\sim}10^{-8}$ to $n_{h}{\sim}10^{-6}cm^{-3}$.
The typical cloud radius is $\sim$100kpc and the 
mass is $\sim10^{8}M_{\odot}$  for the median value  
$n_{h}{\sim}10^{-7}cm^{-3}$.
\vspace{1.0cm}\\
\large
\bf{3. The Ly$\alpha$ Clouds in Hot Galactic Halo}
\vspace{0.3cm}\\
\normalsize
\it
{3.1. Stability Conditions for the Clouds in Galactic Halo}
\rm
\baselineskip 12mm

In addition to the general stability conditions of cold clouds in 
two-phase medium, we must examine additional conditions when  
the Ly$\alpha$ clouds are in a hot galactic halo
( Mo 1994; Mo, Miralda-Escude 1996; Miyahata, Ikeuchi 1995).
 Before we discuss these conditions, we determine $T_{c}$ for a cloud 
embedded in galactic halo. 
Thoul, Weinberg (1996) showed that the equilibrium temperature of the cloud 
irradiated by the UV background radiation varies according to the change of 
both cloud density and the shape and amplitude of the spectrum of radiation.
 Their results for $J_{-21}=1$,
$n_{c}=10^{-2}cm^{-3}$ and $n_{c}=10^{-4}cm^{-3}$
 are shown in Table 2. 
We calculate $T_{c}$ for $J_{-21}=0.01$ in the similar manner as them 
and our results are also shown. 
 We justify later that this range of $n_{c}$ covers the situation 
where we are interested in,
 and show that our final conclution does not change at all 
for the different cases of $n_{c}$, although the cloud properties themselves 
are sensitive to $T_{c}$ (see, eqs (\ref{eqn:rc}) and (\ref{eqn:mc})). 
For this reason, we mainly calculate for the fiducial parameters shown in 
Table 3.  
We now take account of six conditions for a stable cloud in galactic halo.\\
(iv)
When we discuss the Ly$\alpha$ cloud in galactic halo, the timescale 
which characterizes this system would be the dynamical time of galaxy,
$\tau_{halo}$.
 So in this case, 
the stability conditions against the cooling (eq.(\ref{eqn:s1})) and 
evaporation (eq.(\ref{eqn:s3})) are replaced as follows;
\begin{eqnarray}
\tau_{cool}>\tau_{halo},{\hspace{1cm}}\tau_{evap}>\tau_{halo}.
\label{eqn:sg1}
\end{eqnarray}
Here we take $\tau_{halo}$ as the crossing time of the halo 
in our Galaxy that $\tau_{halo}{\sim}R_{h}/v_{rot}{\sim}10^{9}$yr, 
with $R_{h}{\sim}100$kpc and $v_{rot}{\sim}220km~sec^{-1}$.
\\
(v)The cloud must be gravitationally stable, so eq.(\ref{eqn:jeans}) 
holds also in this case.\\
(vi)The clouds should not be tidally disrupted, so that
\begin{eqnarray}
R_{c}<R_{c,tidal}{\sim}
R_{h}(\frac{M_{c}}{M_{gal}})^{1/3}
{\sim}1({\frac{R_{h}}{100kpc}})
({\frac{10^{12}M_{\odot}}{M_{gal}}})^{1/3}
({\frac{M_{c}}{10^{6}M_{\odot}}})^{1/3}
\hspace{0.5cm}kpc,
\end{eqnarray}
where $R_{h}$ and $M_{gal}$ are the typical scale of galactic halo and 
the mass of a galaxy, respectively.
 Together with eqs.({\ref{eqn:rc}}) and eqs.({\ref{eqn:mc}}), 
this criterion is rewritten as follows;
\begin{eqnarray}
\tilde{P}>95.6{\times}T_{4}(\frac{100kpc}{R_{h}})^{3}
(\frac{M_{gal}}{10^{12}M_{\odot}})~~~cm^{-3}K.
\label{eqn:sg2}
\end{eqnarray}
We take $R_{h}{\sim}160h^{-1}kpc{\sim}320kpc$ which is the value 
reported by Lanzetta et.al.(1995), and 
$M_{gal}{\sim}4{\times}10^{11}M_{\odot}$ from the minimal halo model of 
Fish, Tremaine (1991) for the total mass, respectively.
 Using these values, eq.({\ref{eqn:sg2}}) reduces to a simple criterion,
${\log}T_{h}>-{\log}n_{h}+0.54$.
 It is possible that in the extended halo, $M_{gal}$ is larger 
than the value we adopted here (Zaritsky, White 1994), 
but we adopt the above value as conservative one.\\
(vii)The clouds must be stable against the hydrodynamic instability.
 We assume that the 
hydrodynamical instability occurs when the momentum from a cold 
cloud to the hot gas is transferred (Miyahata, Ikeuchi 1995).
So that, we get
\begin{eqnarray}
2n_{c}R_{c}>n_{h}R_{h}.
\end{eqnarray}
Using eq.(\ref{eqn:pc}), we rewrite this as
\begin{eqnarray}
R_{c}>R_{c,crit}=5{\times}10^{2}
(\frac{T_{c}}{10^{4}K})
(\frac{10^{6}K}{T_{h}})
(\frac{R_{h}}{100kpc})
\hspace{0.5cm}pc.
\label{eqn:sg3}
\end{eqnarray}
Our criterion might be a little different from the one for the 
Kelvin-Helmholtz instability, which is one of the most important 
destruction processes when two media having different densities 
are in relative motion (Chandrasekhar 1961; Murray et.al. 1993). 
 But the criterion we examined here would be more generous and realistic.
\\
(viii)The temperature of hot halo of 
which thermal pressure confines a Ly$\alpha$ 
cloud must be roughly equal to the virial temperature of a galaxy.
 Since virial equilibrium for the galaxy says
\begin{eqnarray}
0.6\frac{GM_{gal}}{R_{h}}=3\frac{kT_{h}}{m_{p}},
\end{eqnarray}
the virial temperature of halo is written as,
\begin{eqnarray}
T_{h}=1.70{\times}10^{7}(\frac{M_{gal}}{10^{12}M_{\odot}})
(\frac{100kpc}{R_{h}})~~K.
\label{eqn:sg4}
\end{eqnarray}
By substituting the above value, this relation reduces to $T_{h}{\sim}
2{\times}10^{6}$K.
 In addition to this condition, the luminosity of hot extended halo 
inferred from X-ray observation (Sarazin 1996) might also constrain 
our model when we consider more realistic density profile of galactic 
halo. We discuss this point later.

In this paper, we assume that the Ly$\alpha$ cloud is irradiated only by 
the diffuse UV background radiation. 
But in more realistic situations when such a small cloud is in galactic 
halo, additional sources of UV radiation such as young stars or AGN 
(Kang,et.al. 1990) may be taken account of.
Recently, Norman, Spaans(1997) and Spaans, Norman (1997) studied a 
multi-component protogalaxy model in the presence of UV radiation both 
from background source and young stars in galaxy, as a model for 
protogalactic disk and dwarf galaxy, respectively.   
Here we neglect such sources because we do not have 
any definite data  about the amount 
of the radiation from such sources, 
although it is noted that from eq.(\ref{eqn:rc}) 
we expect that the evaporation condition 
is effectively relaxed under additional UV sources.   
\vspace{0.5cm}\\
\baselineskip 12mm
\it
{3.2. Results}
\rm
\baselineskip 12mm

In Figure 3a (Model 3) and Figure 4 (Model 4), we show the stability 
condition for a cloud in galactic halo except for those against 
hydrodynamic instability, which is less important in our case.
 These figures show that in both cases of z=2.5 and z=0.5, 
the condition that a cold cloud is stable against evaporation 
as well as tidal disruption is very narrowly limited. 

 As a result, there remains a little allowed region for 
$n_{h}=10^{-4.8}cm^{-3}$ and $T_{h}=10^{5.5}$K at z=2.5.
 But such a low temperature gas is hardly expected in typical galactic halo.
 So, it might be concluded that the pressure-confined Ly$\alpha$ 
cloud can not survive in galactic halo at z=2.5.
 In the most noteworthy case, z=0.5, there is 
no allowed region for a cloud, too. 
 So we conclude that pressure-confined, spherical Ly$\alpha$ clouds with  
typical column density $N_{HI}=10^{14}cm^{-2}$ cannot survive
 in galactic halo, in general.
In Figure 3b, we show the results for a cloud which has two orders of 
magnitude lower density and a little higher temperature (Table 2). 
Comparing Figure 3a and Figure 3b, we can recognize that variation of 
cloud temperature and density does not change our conclusion concerning the 
stability of cloud.

The same analyses have been done for the same redshift but with different 
column density of a cloud, and results are shown in Figure 5a and Figure 5b.
 For a smaller cloud of $N_{HI}=10^{12}cm^{-2}$, the evaporation condition 
becomes severer and a cloud quickly disappears, as is expected.
 In contrast to this, for a cloud of higher column density 
$N_{HI}=10^{17}cm^{-2}$, which is hard against evaporation, there remains 
allowed region where typical quantites of a cloud are 
$R_{c}{\sim}$10kpc and $M_{c}{\sim}10^{6.5}M_{\odot}$ 
and those of galactic halo are 
$n_{h}{\sim}10^{-6}cm^{-3}$ and $T_{h}=10^{6.3}$K.
 Such a cloud is not destructed by hydrodynamical instability and 
$T_{h}$ is comparable to the virial temperature of galactic halo.
 For a case of $N_{HI}=10^{16}cm^{-2}$, the cloud survival is marginal 
and we suppose that this column density is the critical one for the cloud 
to survive in galactic halo.
 Although, 
pressure confined clouds in galactic halo fail to reproduce 
the observed quantities 
of Ly$\alpha$ clouds, it is noted that this model may be 
 still valid for a model of metallic absorption line systems 
which associate with galaxy (Steidel 1995).
\vspace{0.5cm}\\
\large
\bf
4.Summary and Discussion
\rm
\normalsize
\baselineskip 12mm

Motivated by the recent observations (Lanzetta et.al.1995,1997a) 
which report that
 most of the nearby(z$<$1) Ly$\alpha$ clouds are associated with  
extended gaseous envelopes of galaxies with impact parameter 
${\sim}160h^{-1}$kpc,
 we examined the conditions for stable Ly$\alpha$ clouds within the context of 
 two-component gaseous medium.
 In the case for protogalaxy model, it is expected that the association of the 
Ly$\alpha$ clouds with galaxies is naturally expected.
 We determined the physical quantities of both the cold cloud 
and the hot ambient gas 
on the basis of various stability conditions.
 For simplicity, we assumed that the cold cloud confined by 
the pressure of 
 the hot ambient medium is spherical, homogeneous, and in thermal and 
ionization equilibria irradiated by the diffuse UV background radiation.

We calculated  physical quantities of the pressure-confined Ly$\alpha$ cloud 
in high redshift(z=2.5) and low redshift(z=0.5).
One is in intergalactic medium and the other is in galactic halo.
 In the former case, the upper mass bound of the cloud is given by the 
stability condition against Jeans instability, while the lower mass bound
 of it is constrained by the evaporation condition, respectively.
 In the latter case, in addition to the above-mentioned constraints 
the stability conditions for 
the cloud with respect to the tidal disruption gives a 
severe upper bound of the cloud radius, while the hydrodynamical  
stability condition gives the lower bound of it.

As for a pressure-confined spherical cloud in intergalactic medium, 
our model is consistent to some of the observed properties of Ly$\alpha$ 
clouds.
 So such a model is still valid for Ly$\alpha$ clouds detected in 
intergalactic medium (Shull et.al. 1995) together with an 
alternative model for a Ly$\alpha$ cloud 
 which has little or no correlation with each other and with galaxies
(Ikeuchi 1986, Rees 1986, 
Miralda-Escude et.al. 1996; Cen, Simcoe 1997; Zhang et.al. 1997; 
Weinberg et.al. 1997 and references therein).

For a cloud in galactic halo, a smaller cloud is quickly 
evaporated due to heat 
conduction from ambient hot medium and a larger cloud is tidally disrupted 
by galactic halo potential.
 As a result, it is concluded that for both cases of z=2.5 and z=0.5 
pressure-confined Ly$\alpha$ cloud with typical column density 
$N_{HI}=10^{14}cm^{-2}$ cannot survive in galactic halo, 
although a cloud having higher column density $N_{HI}=10^{17}cm^{-2}$ 
can do there in stable.

In our simple treatment, several factors are neglected. 
Do they change our conclusion? 

First, the thermal conductivity may be fairly reduced if the 
 magnetic field exists(Pistinner et.al. 1996 and references therein), 
and a severe lower bound for the mass of a Ly$\alpha$ cloud may be relaxed 
as well as the suppression of hydrodynamical instability. 
 If so, the pressure-confined Ly$\alpha$ cloud of $N_{HI}=10^{14}cm^{-2}$ 
 may exist in galactic halo, but we have few knowledge concerning 
the magnetic field to an extent of ${\sim}$ 100kpc of galaxy so that 
we cannot analyze quantitatively this effect on cloud properties.

Second, we do not take into account of the UV radiation from young stars 
in galaxy(Norman, Spaans 1997; Spaans, Norman 1997). As we mentioned in the 
previous section, this UV component effectively relaxes the 
evaporation condition.
 But at the same time, evaporation time scale we estimate becomes shorter 
due to electrons in heavy elements synthesized in newly formed stars.
 Both of these effects are difficult to include quantitatively in 
this stage.

Third, we do not take into account the hierarchical clustering 
picture of galaxy formation 
(Cole et.al. 1994, Kauffmann et.al. 1993). 
 If galaxies are formed in this way, 
we have to replace $\tau_{halo}$ in eqs.({\ref{eqn:sg1}}) by 
a typical merging timescale of halos, ${\Delta}t_{merge}$, 
and to consider various stability conditions for a cloud at each time 
when galaxies merged and resultant halo heated up. 
 It is too complicated for us to follow this history at present, 
but we can simply indicate that our conclusion does not change at all 
as far as ${\Delta}t_{merge}$ is in order of $10^{8}$year.

Finally, we speculate how the alternative model for the Ly$\alpha$ cloud 
may be, from the viewpoint of 
 our conclusion that pressure-confined, spherical Ly$\alpha$ cloud 
with $N_{HI}=10^{14}cm^{-2}$ cannot survive in galactic halo. 
 The other spherical cloud models may be classified to two categories.
 One is the cloud confined by the gravity 
of CDM and the other is a ram pressure-confined cloud which will finally 
 collapse in a longer timescale than a free-fall time of a cloud.

When the cold cloud is gravitationally confined 
by the CDM potential, the stability 
condition against the tidal disruption may be relaxed,   
 hydrodynamical instability is greatly suppressed, 
and the properties of hot halo are not constrained through cooling and 
evaporation condition. 
Therefore, it is highly probable that 
the non-evolving component of Ly$\alpha$ forest in associated with 
galaxies may be the clouds confined by the CDM, 
and the evolving one in intergalactic medium may 
be the pressure confined clouds.

Ram pressure-confined, cloud model would be emphasized as follows.
 In this paper we assume that the extended gaseous halo is isothermal
($n_h{\sim}r^{-2}$) which is expected in our Galaxy. 
 Using this relation and eq. (\ref{eqn:nh1r})
, we can predict a remarkable feature for the Ly$\alpha$ clouds.
 In the single halo, higher $N_{HI}$ systems exist near the center of the halo 
and lower $N_{HI}$ systems exist in the outer region of it. 
 The correlation between the equivalent widths of Ly$\alpha$ absorption lines 
and the impact parameters from nearby galaxies indicated by Lanzetta et.al.
(1995, 1997) may reflect this distribution law.
 So in our next step to examine the physical properties of 
pressure-confined Ly$\alpha$ clouds associated with galaxy, 
 it may be natural to analyze the properties of a cloud for  
various positions of a single halo and for various halos.
 For example, at a different distance from galactic center, 
the stable conditions against evaporation and tidal disruption 
for a cloud as well as the typical column density $N_{HI}$ of a cloud 
are also different.
After such an estimate has been done, we will compare the results 
with recent data for Ly$\alpha$ clouds which associate with 
galaxies,---strong correlation of Ly$\alpha$ absorption equivalent width, 
galaxy impact parameter, and galaxy B-band luminosity(Chen et.al. 1997; 
Lanzetta et.al. 1997b). 
 From those analyses, we will also be able to predict physical quantities  
 of hot halo which posesses Ly$\alpha$ clouds 
in relation to X-ray observation(Sarazin 1996)
or extreme-UV observation of galactic halo. 
\vspace{2cm}\\
The authors would like to thank Professors M.Sasaki and N.Gouda for valuable 
discussion, and
K.M. is grateful to Dr.M.Shibata for critical reading of the manuscript 
and his comments. 
She also wishes to thank Professor M.Strauss, Dr.I.Murakami, Dr.K.Denda, 
and Dr.Y.Fujita for their advice. 
This paper is partly based on research support by the Research Fellow of 
the Japan Society for the Promotion of Science No.2006.
\\ 
\newpage
\large
\begin{center}
\bf{References}
\end{center}

\normalsize
\baselineskip 12mm
\noindent
Bahcall J.N., Spitzer L.Jr.  1969, ApJ 156, L63\\
Balbus S A., McKee C.F. 1982\\
Bechtold J.,et.al. 1994, ApJ 437, L83\\
Binney J., Tremaine S. 1987, {\it Galactic Dynamics} (Princeton)\\
Black J.H. 1981, MNRAS 197, 553\\
Bowen D.V., et.al. 1996, ApJ 464, 141\\
Brun V.L., et.al. 1996, Astr.Ap. 306, 691\\
Cen R., Simcoe R.A. 1997, ApJ 483, 8\\
Chandrasekhar S. 1961, {\it{Hydrodynamic and Hydromagnetic Stability}}
 (Clarendon Press, Oxford)\\
Chen H-W., et.al. 1997, to appear in the proceeding of 13th 
IAP colloquim, {\it{Structure and Evolution of the Intergalactic Medium 
from QSO Absorption Line Systems}} (astro-ph/9709173)\\
Chiba M., Nath B.B. 1997, ApJ 483, 638\\
Cole S., et.al. 1994, MNRAS 271 781\\
Dinshaw N., et.al. 1994, ApJ 437, L87\\
Fall S.M., Rees M.J. 1985, ApJ 298, 18\\
Ferrarra A., Field G.B. 1994, ApJ 423, 665\\
Field G.B. 1965, ApJ 142 531\\
Field B.D. et.al. 1997, ApJ 483 625\\
Fish M., Tremaine S. 1991 ARAA 29, 409\\
Fukugita M. et.al. 1996, Nature 381,489\\
Ikeuchi S. 1986, ApSS 118, 509\\
Ikeuchi S., Ostriker J.P. 1986, ApJ 301, 522\\
Ikeuchi S., Norman C.A. 1991, ApJ 375 479\\
Ikeuchi S. 1995, in the proceeding of
 {\it Cosmological Constant and the Evolution of the Universe,},
 eds. Sato K., et.al. (Univ.acad.press), 119\\
Kang H.,et.al. 1990, ApJ 363, 488\\
Kauffmann G., et. al. 1993 MNRAS 264, 201\\
Lanzetta K.M., et.al. 1995, ApJ 442, 538\\
Lanzetta K.M., et.al. 1997a, to appear in the proceeding of 
{\it{the 18th Texas Symposium on Relativistic Astrophysics}} 
, eds. Olinto A., et.al. \\
Lanzetta K.M., et.al.
1997b, to appear in the proceeding of 13th 
IAP colloquim, {\it{Structure and Evolution of the Intergalactic Medium 
from QSO Absorption Line Systems}} (astro-ph/9709168)\\
Miralda-Escude J., et.al. 1996, ApJ 471,582\\
Miyahata K., Ikeuchi,S. 1995, PASJ 47, L37\\
Mo H.J., 1994, MNRAS 269, L49\\
Mo H.J., Miralda-Escude J., 1996, ApJ 469, 589\\
Morris S.L., to appear in the proceeding of {\it{GHRS Science Symposium}}
(astro-ph/9610225)\\
Murray S.D., et.al. 1993, ApJ 407 588\\
Murray S.D., Lin D.N.C., 1993, in the proceeding of 
{\it{the Globular Cluster-Galaxy Connection}}, 
eds. Smith,G.H., Brodie,J.P. (A.S.P. Conf.Ser. vol 48)\\
Norman C.A. 1994 in {\it{Galaxy Formation}} eds. Silk J., Vittorio N. 
(North-Holland), 283\\
Norman C.A., Spaans M. 1997 ApJ 480 145\\
Ostriker J.P., Ikeuchi S. 1983, ApJ 268, L63\\
Pistinner S. et.al. 1996, ApJ 467,  162\\
Rees M. 1986, MNRAS 218, L25\\
Rees M., Ostriker J.P. 1977, MNRAS 179, 541\\
Sarazin C.S. 1996, 
to appear in the proceeding of {\it {The Nature of Elliptical Galaxies}},
 eds. Arnaboldi G.S. et al. (San Francisco: Publ.Astr.Soc.Pacific), (astro-ph/9612054)\\
Savaglio S., Webb J. 1995, in the proceeding of 
{\it QSO Absorption Lines}, eds. Meylan G. (Springer), 96\\
Shull J.M. et.al. 1996, AJ 111, 72\\
Shull J.M. 1997, to appear in the proceeding of 13th 
IAP colloquim, {\it{Structure and Evolution of the Intergalactic Medium 
from QSO Absorption Line Systems}} 
astro-ph/9709110\\
Songaila A., Cowie L.L. 1996, AJ 112, 335\\
Spaans M., Norman C.A., 1997, ApJ 483, 87\\
Spitzer L.Jr. 1978,
{\it Physical Processes in the Intersteller Medium} 
(Wiley Interscience,New York)\\
Spitzer L.Jr. 1990, ARAA 28, 71\\
Spitzer L.Jr., Ostriker J.P. 1997 in {\it{Dreams, Stars, and Electrons}}
 (Princeton Univ. Press)\\
Steidel C.C. 1995, in the proceeding of {\it QSO Absorption Lines},
eds. Meylan G. (Springer),139\\
Stocke et.al. 1995, ApJ 451, 24\\
Thoul A.A., Weinberg D.H. 1996, ApJ 465 608\\
Tytler D. 1995, in the proceeding of {\it QSO Absorption Lines},
eds. Meylan G. (Springer),289\\
Ulmer A. 1996, ApJ 473, 110\\
Weinberg et.al. 1997, to appear in {\it{Origins}}, eds. Shull J.M. et.al. 
(ASP Conf. Ser.), (astro-ph/9708213)\\
Weymann R. 1995, in the proceeding of {\it QSO Absorption Lines},
eds. Meylan G. (Springer), 3
Zaritsky D., White S.D.M. 1994 ApJ 435 599\\
Zhang Y. et.al., 1997, to appear in ApJ (astro-ph/9706087)\\ 
\newpage
\begin{center}
\large
\bf{Figure Captions}
\end{center}
\bf{Fig. 1a}\\
\rm
\normalsize
The three conditions presented in
$\S$2.2.  are shown in 
the plane of density and temperature of the hot ambient gas
 of two-component system at z=2.5 (Model 1).
This situation essentially 
corresponds to Ly$\alpha$ cloud in hot intergalactic medium.
 The hatched sides are forbidden. 
 As we noted in the text, the right below, left below and right upper 
region in this figure are not allowed due to the 
cooling, gravitationally unstable, and evaporation conditions, respectively.  
 For comparison, a constant cloud radius and a constant cloud 
mass which discussed in $\S$2.1 are shown for the cases of 
$R_{c}=$100,10 and 1kpc (short-dashed line) and of 
$M_{c}=10^{8}$, $10^{6}$ and $10^{4}M_{\odot}$ (dot-dashed line)
 from left to right, respectively.
\vspace{1cm}\\
\bf{Fig. 1b}\\
\rm
\normalsize
The allowed mass ranges of cold clouds is shown with respect to 
 the pressure of the hot phase for the case of $T_{4}=J_{-21}=1$.
 The evaporation conditions are shown for three cases of $n_{h}=10^{-5}$,
$10^{-6}$ and $10^{-7}cm^{-3}$.
 Note that roughly speaking, the upper and lower mass bound of the cloud is 
constrained by the stability condition against the Jeans instability 
and evaporation, respectively. 
The allowed region is indicated in the trapezoid region left below for the 
three cases of the density of the hot halo (solid line). 
For comparison, a constant cloud radius is shown for the case of 
$R_{c}=$100, 10 and 1kpc from top to bottom
 (short-dashed line), and a constant neutral hydrogen 
column density is also plotted (dash-dotted line).
\vspace{1cm}\\
\bf{Fig. 2a}\\
\normalsize
The same as in Figure1(a) 
but for the case of a cloud in two-component system at z=0.5
 (Model 2).
 A constant cloud radius and a constant cloud mass are shown for the cases of 
$R_{c}=$100kpc (short-dashed line) and of 
$M_{c}=10^{8}M_{\odot}$ (dot-dashed line).
\vspace{1cm}\\
\bf{Fig. 2b}\\
\rm
The same as in Figure1(b), but for the case of cloud at z=0.5.
 The evaporation conditions are shown for three cases of $n_{h}=10^{-6}$,
$10^{-7}$, and $10^{-8}cm^{-3}$.
 Lines of constant $R_{c}$ and $N_{HI}$ are shown for the same parameters as 
 in Figure1(b).
\vspace{1cm}\\
\bf{Fig. 3a}\\
\rm
\normalsize
The conditions presented in
$\S$3.1 are shown in 
the plane of density and temperature of the hot halo 
 of two-component protogalaxy at z=2.5 (Model 3).
 For such a cloud, the stability against the tidal disruption   
gives a lower bound on a pressure $\tilde{P}$, which is shown 
by a dotted line.
 Horizontal dotted line shows the virial temperature of typical galaxy.
 Lines of constant $R_{c}$ and $M_{c}$ are shown for 1kpc and 
$10^{4}M_{\odot}$.
\vspace{1cm}\\
\bf{Fig. 3b}\\
\rm
\normalsize
The same as in Figure 3a but for a cloud of $T_{4}$=6.3. 
 Lines of constant $R_{c}$ and $M_{c}$ are shown for 1kpc and
$10^{6}M_{\odot}$.
\\
\bf{Fig. 4}\\
\rm
\normalsize
The same as in Figure3(a) but for a cloud in a hot galactic halo at z=0.5
(Model 4).
This figure apparently shows that there is no allowed region.
\vspace{1cm}\\
\bf{Fig. 5a}\\
\rm
\normalsize
The same as in Figure 4 but for the case of $N_{HI}=10^{12}cm^{-2}$ (Model 5).
\vspace{1cm}\\
\bf{Fig. 5b}\\
\rm
\normalsize
The same as in Figure 4 but for the case of $N_{HI}=10^{17}cm^{-2}$ (Model 6).
 Lines of constant $R_{c}$ and $M_{c}$ are shown for 10, 1kpc and
$10^{6}$,$10^{4}M_{\odot}$ from left to right, respectively.
\begin{center}
Table1.~Symbols and their meanings
\vspace{0.5cm}\\
\begin{tabular}{ccc} \hline
symbol & unit & meaning\\
\hline
\hline
z & & redshift\\
$\Omega$ & & density parameter \\
$H_{0}$ & $Mpckm^{-1}sec^{-1}$ & Hubble constant \\
$n_{h}$ & $cm^{-3}$ & hot phase density \\
$T_{h}$ & K & hot phase temperature \\
$n_{c}$ & $cm^{-3}$ & cloud density \\
$T_{c}$ & K & cloud temperature \\
$T_{4}$ & $10^{4}$K & normarized cloud temperature\\
$R_{c}$ & pc & cloud radius \\
$M_{c}$ & $M_{\odot}$ & cloud mass \\
$R_{h}$ & pc & typical scale of galaxy \\
$M_{gal}$ & $M_{\odot}$ & total mass of galaxy \\
$N_{HI}$ & $cm^{-2}$ & HI column dencity of cloud\\
$N_{14}$ & $10^{14}cm^{-2}$ & normalized HI column density\\
$J_{-21}$ & $10^{-21}ergcm^{-2}s^{-1}Hz^{-1}str^{-1}$ & mean UV intensity\\
$\tau_{H}$ & sec & Hubble time \\
$\tau_{cool}$ & sec & cooling time \\
$\tau_{evp}$ & sec & evaporation time \\
$\tau_{halo}$ & sec & crossing time of halo \\
\hline
\end{tabular}
\vspace{2cm}\\
Table2.~Physical Processes and adopted values of Parameters
\vspace{0.5cm}\\
\begin{tabular}{ccc} \hline
parameters  & adopted value & references\\
\hline
\hline
z & 0.5, 2.5 & --- \\
$\Omega$ & 1 & --- \\
$H_{0}$ & 50 & --- \\
$J_{-21}$ & $10^{-2}$(z=0.5), 1(z=2.5) & Savaglio$\&$Webb\\
$N_{14}$   & 1,$10^{-2}$,$10^{3}$ & ---\\
$T_{4}$(cloud in IGM) & 3.0 & Ikeuchi$\&$Ostriker\\
 &2.5($J_{-21}=1,n_{c}=10^{-2}$) & \\
\multicolumn{1}{c}{\raisebox{-1.5ex}[0pt]{$T_{4}$(cloud in halo)}} & 
 6.3($J_{-21}=1,n_{c}=10^{-4}$) &
\multicolumn{1}{c}{\raisebox{+1.5ex}[0pt]{Thoul$\&$Weinberg}} \\
 &1.2($J_{-21}=10^{-2},n_{c}=10^{-2}$) & 
\multicolumn{1}{c}{\raisebox{-1.5ex}[0pt]{this work}} \\
 &2.3($J_{-21}=10^{-2},n_{c}=10^{-4}$) & \\
\hline
\end{tabular}
\vspace{2cm}\\
Table3.~Model parameters in our calculation
\vspace{0.5cm}\\
\begin{tabular}{cccccc} \hline
model number & cloud situation & z ($J_{-21}$) & $N_{14}$ & $T_{4}$\\
\hline
\hline
Model 1 & IGM & 2.5~(1) & 1 &  3\\
Model 2 & IGM & 0.5~($10^{-2}$) & 1 & 3 \\
\cline{2-5}
Model 3 & galactic halo & 2.5~(1) & 1 & 2.5  \\
Model 4 & galactic halo & 0.5~($10^{-2}$) & 1 & 1.2 \\
Model 5 & galactic halo & 0.5~($10^{-2}$) & $10^{-2}$ & 1.2 \\
Model 6 & galactic halo & 0.5~($10^{-2}$) & $10^{3}$ & 2.3 \\
\hline
\end{tabular}
\vspace{2cm}\\
Table4.~Summary of Results;~
Allowed physical quantities for the Ly$\alpha$ cloud
\vspace{0.5cm}\\
\begin{tabular}{ccc} \hline
model number & log$R_{c}(pc)$ & log$M_{c}(M_{\odot})$ \\
\hline
\hline
Model 1& 3.0${\sim}$5.4 & 4.1${\sim}$9.2 \\
Model 2& 3.6${\sim}$6.2 & 6.8${\sim}$9.3 \\
\cline{2-3}
(Model 3)& (2.9${\sim}3.4)$ & (4.0${\sim}$4.7)\\
Model 4&${\times}$&${\times}$\\
Model 5&${\times}$&${\times}$\\
Model 6& 3.8${\sim}$ 4.2 & 6.4${\sim}$6.9 \\
(Model 6)& (2.6${\sim}$4.6) & (3.9${\sim}$7.9) \\
\hline
\end{tabular}
\vspace{0.5cm}\\
(${\sim}$): results for $T_{h}=10^{5.5}$K not $T_{h}=10^{6.3}$K \\
${\times}$: no allowed region
\vspace{2cm}
\end{center}
}
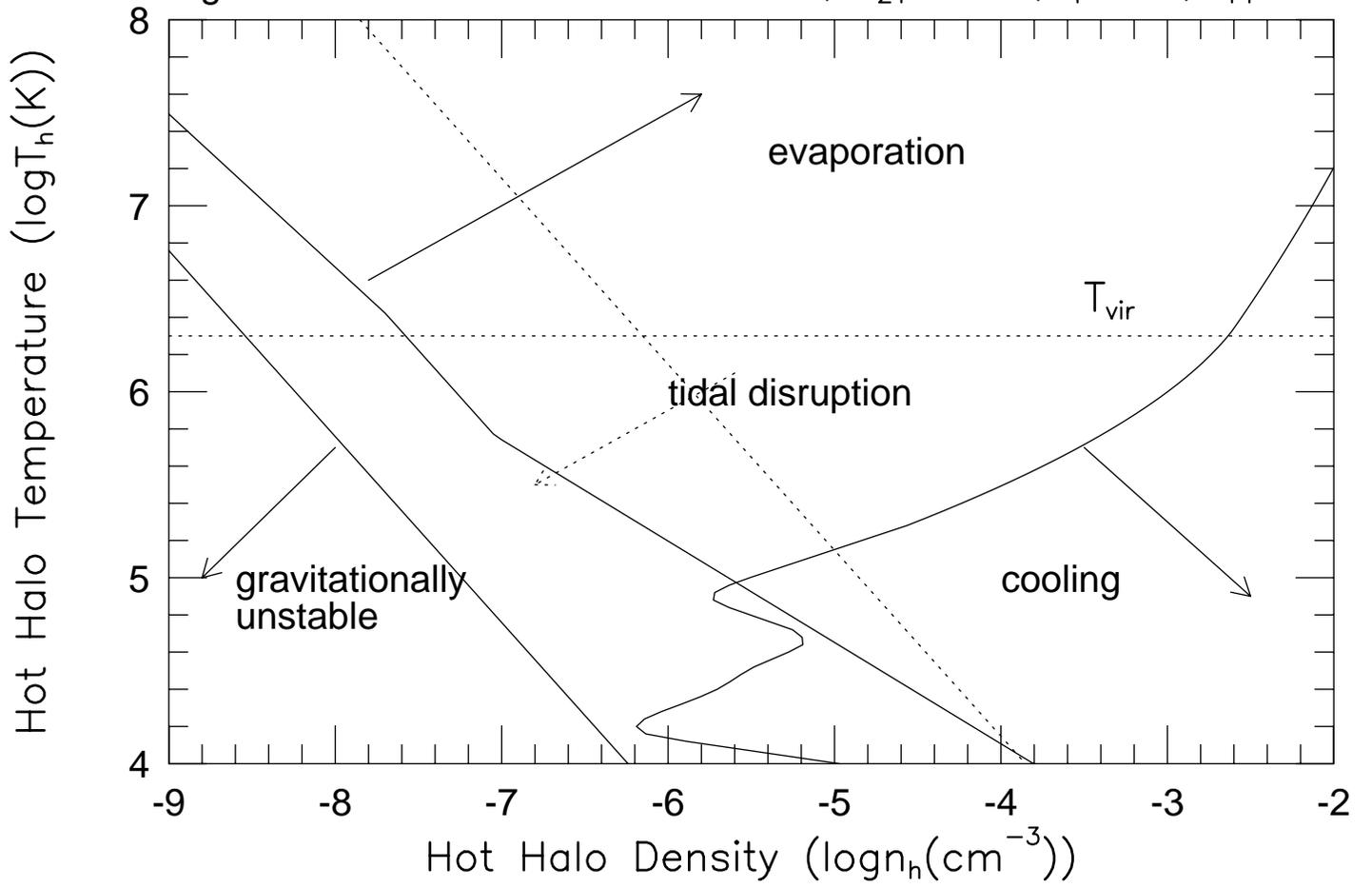
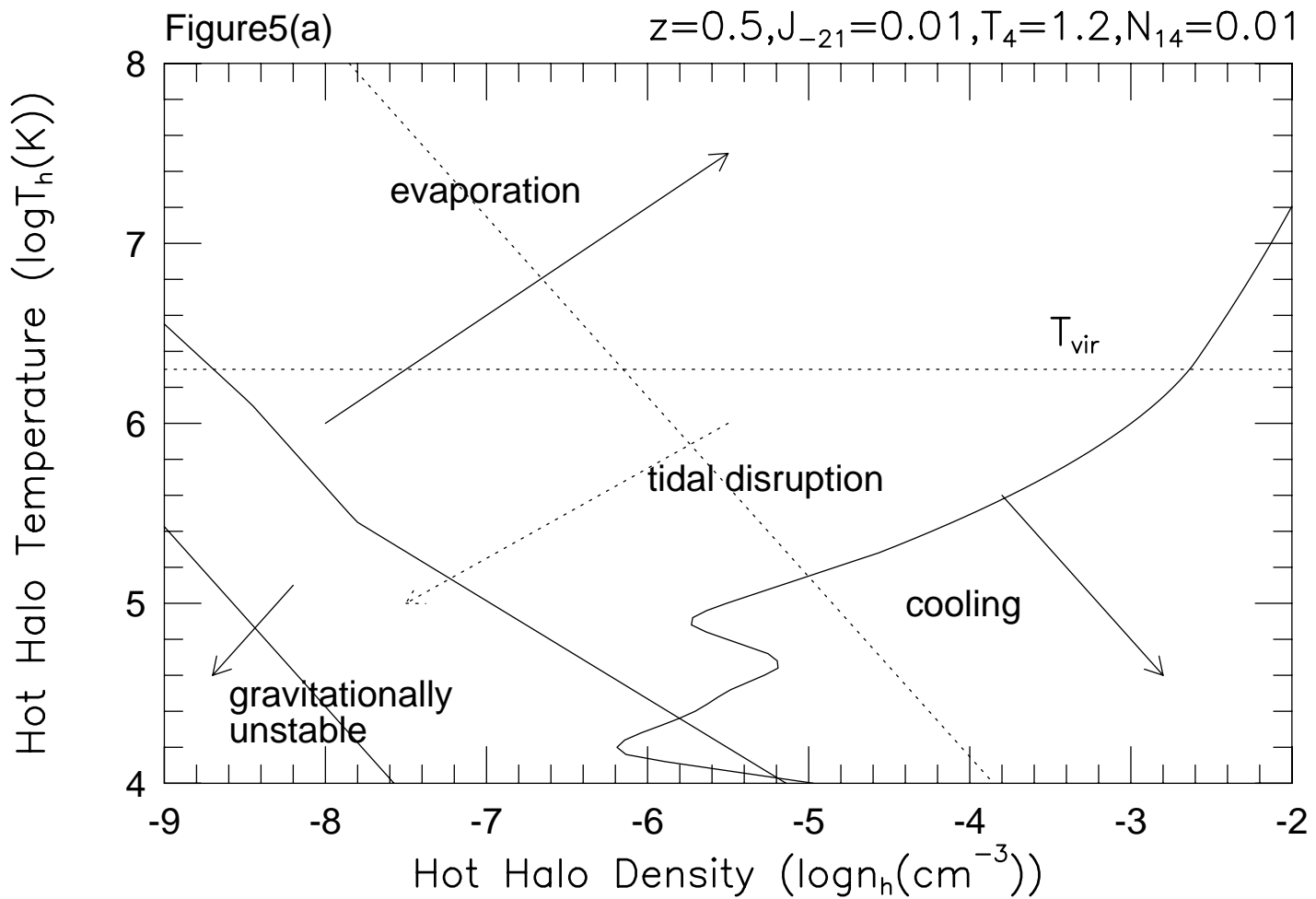
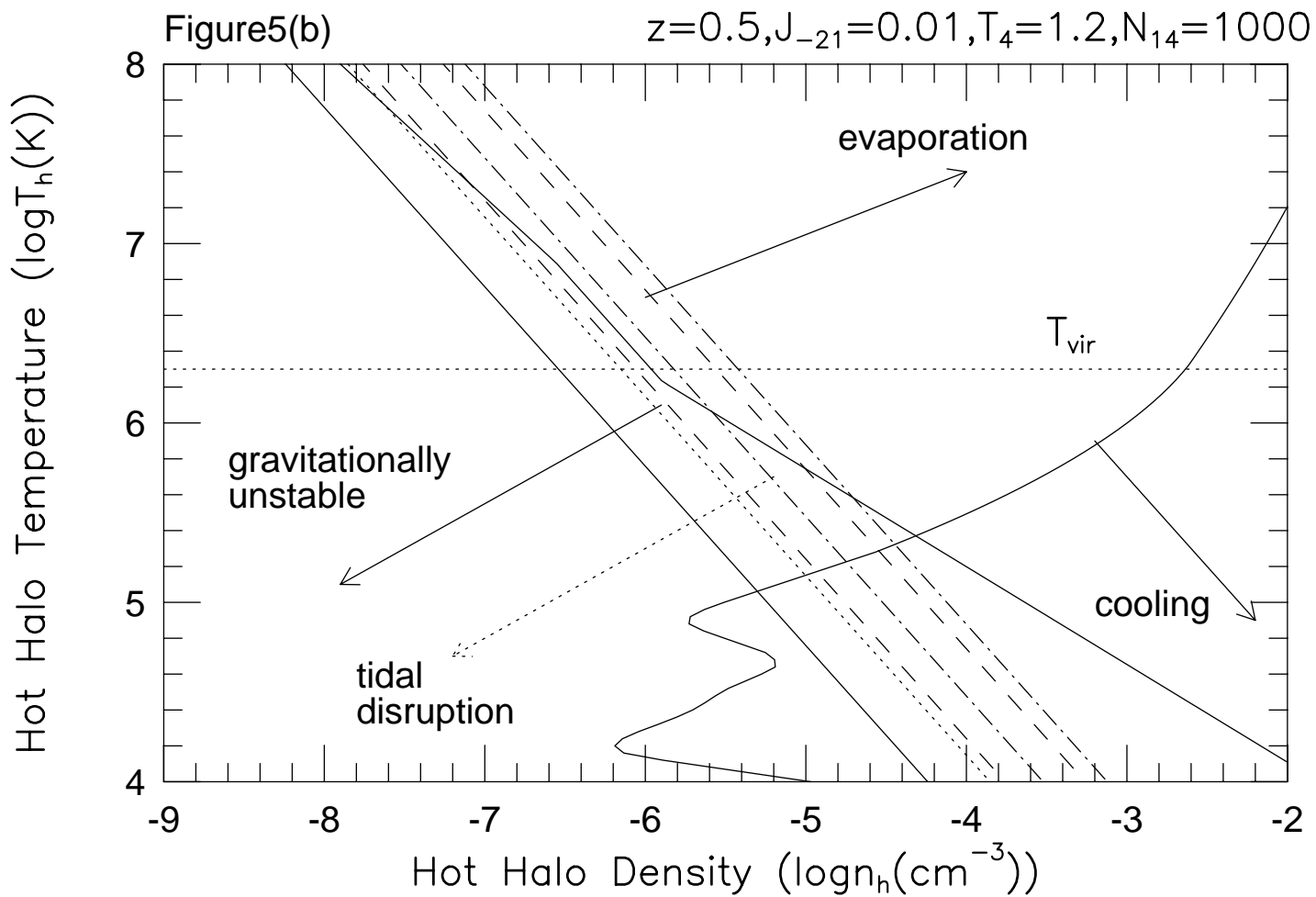
\end{document}